\documentclass[a4paper,11pt]{article}
\usepackage{amsmath,amssymb}

\usepackage[usenames]{color}
\usepackage{graphicx}
\usepackage{hyperref}
\usepackage{cite}
\usepackage{url}

\begin{document}

\begin{center}
{\Large \bf Molecular Dynamics Analysis of Biomolecular Systems Including Nucleic Acids}
\medskip

Takeru Kameda\textsuperscript{1,2,3},
Akinori Awazu\textsuperscript{4},
Yuichi Togashi\textsuperscript{1,4,5,6$*$}
\end{center}
\textbf{1} {College of Life Sciences, Ritsumeikan University, Kusatsu, Shiga 525-8577, Japan} \\
\textbf{2} {Graduate School of Science, Hiroshima University, Higashi-Hiroshima, Hiroshima 739-8526, Japan} \\
\textbf{3} {RIKEN Center for Biosystems Dynamics Research (BDR), Wako, Saitama 351-0198, Japan} \\
\textbf{4} {Research Center for the Mathematics on Chromatin Live Dynamics (RcMcD), Graduate School of Integrated Sciences for Life, Hiroshima University, Higashi-Hiroshima, Hiroshima 739-8526, Japan} \\
\textbf{5} {RIKEN Center for Biosystems Dynamics Research (BDR), Higashi-Hiroshima, Hiroshima 739-0046, Japan} \\
\textbf{6} {Cybermedia Center, Osaka University, Ibaraki, Osaka 567-0047, Japan} \\
$*$ togashi@fc.ritsumei.ac.jp


\section*{Abstract}

Along with recent progress in structural biology and genome biology, structural dynamics of molecular systems including nucleic acids has attracted attention in the context of gene regulation. Structure-function relationship is an important topic, where physicochemical properties of nucleotides are important as well as those of amino acids in proteins. Simulation is a useful tool for the analysis of molecular dynamics in detail, complementary to experiments. However, molecular simulation of nucleic acids has less developed than that of proteins yet, partly due to the physical nature of nucleic acids. In this review, we briefly describe the current situation and future directions of the field, hopefully as a guide to collaboration between experiment and computation.

\section*{Significance}

DNA carries genomic information, and RNA serves as a catalyst or regulator as well as a temporal copy of the information.
These nucleic acids are not merely information media but also parts of the information processing machinery,
and hence their physical properties are important for the operation of the system.
However, compared to proteins and lipids, molecular simulation of nucleic acids has its intrinsic difficulty
and is still much to be improved.
We overview recent progress in this field and discuss future directions.

\section{\label{sec:intro} Introduction}

Biological systems are characterized by a huge variety of macromolecules.
The function of these macromolecules is inseparable from the structure; particularly, the operation of molecular machines such as enzymes is coupled with their conformational changes, analogous to macroscopic machines.
Structural biology from the viewpoint of physics and chemistry has thus expanded its scope of application and attracted broad attention in life sciences.
X-ray crystallography \cite{boutet2012high,spence2012x,martin2016serial} and nuclear magnetic resonance (NMR) spectroscopy \cite{kay2011nmr,prestegard1998new} have determined many protein structures.
Recently, cryo-electron microscopy (cryoEM) has become a powerful tool for large complexes \cite{topf2008protein,yip2020atomic,shi2013three,nogales2016development,ho2018rapid,raunser2017cryo}.
Analysis using these methods is driving the progress in structural biology \cite{branden2012introduction}.
Protein structure and function (e.g. enzyme activity) is unveiled in parallel \cite{sadowski2009sequence};
which will further proceed in the future, as there remain a lot of proteins without known structure.

Besides, these methods have been applied to other biomolecules.
Since the discovery of the double-helix structure of DNA by Watson and Crick \cite{crick1953molecular}, nucleic acids have been also targeted \cite{sim2012modeling,blackburn2006nucleic,mccammon1988dynamics}.
As well as the amino acid sequence in proteins, the nucleotide sequence affects the structure and physical properties of nucleic acids \cite{rief1999sequence,ma2016anisotropy}.
Of course, complexes of nucleic acids and proteins play crucial roles in the cell.
Nucleosomes, the constituent units of chromatin, are of particular importance, and their structure has been determined \cite{davey2002solvent,vasudevan2010crystal,tan2011nucleosome}.
Efforts to elucidate the ties between the nucleosome structure and dynamics and the transcription regulation mechanism are now acknowledged in the emerging field ``chromatin biology'' \cite{luger2005nucleosome,dai2020evolving,hihara2012local}.

Although these experiments have contributed to a huge number of structures, information on the dynamics is limited \cite{rao2010protein}.
Analysis of the dynamics is particularly important if the target molecule shows large conformational change or secondary structure transition.
However, crystallography and cryoEM can provide only single static structures in most cases, and the spatiotemporal resolution of NMR is not very high.
Molecular simulation has thus become a complement to these experimental methods \cite{frenkel2001understanding,kofke2004getting,bonomi2019biomolecular,huggins2019biomolecular}.
In this review, we briefly introduce the basics of molecular dynamics (MD) simulations
and then showcase recent advances in MD studies of proteins and nucleic acids.
Finally, we discuss the current situation and future directions of molecular modeling and simulation for nucleosome and chromatin.

\section{\label{sec:simulation} Molecular Dynamics Simulation}

\subsection{\label{sec:general} Overview of Molecular Simulation}

Molecular simulation namely reproduces the behavior of molecules {\em in silico} \cite{frenkel2001understanding,kofke2004getting,bonomi2019biomolecular,huggins2019biomolecular}.
Simulation always requires a model, generally represented by governing equation or algorithm.
Then, the state variables of the model are (approximately) calculated on computers,
e.g. by numerically solving the differential equation for the time evolution of the system, 
or by the Monte Carlo method using random numbers for sampling structures or reproducing stochastic behavior.

In MD simulation, for example, the target molecule is represented by particles (mass points); e.g. in all-atom models, each atom corresponds to a particle.
Assuming that the force on each particle depends only on the configuration of the particles (force field) and the motion obeys classical (Newtonian) mechanics, the equation of motion for each particle is numerically solved \cite{karplus2002molecular,karplus1990molecular}.
Of course, the model should be chosen according to the target system and property \cite{bao2002mechanics}.
If chemical reaction may matter, i.e. the electron transfer is involved, quantum mechanics represented by the Schr\"{o}dinger equation should be considered for the electronic structure \cite{gogonea2001new,stock2009classical,donchev2005quantum}.
However, the computational cost for {\em ab initio} quantum chemistry methods is generally too large to track motion (i.e. calculation repeated for many time steps; {\em ab initio} or first-principle MD) of macromolecules such as proteins \cite{dunning1977gaussian}.
Hence, classical MD simulation is widely used when applicable \cite{hansson2002molecular,karplus2002molecular,karplus1990molecular,karplus2005molecular} (or quantum mechanics is combined only where necessary e.g. reaction center; QM/MM method).

In classical MD simulation, as mentioned above, only Newtonian mechanics is considered \cite{binder2004molecular,li2005basic,paterlini1998constant,tuckerman1992reversible}.
Using all-atom models, nevertheless, the accessible time scale is only nanoseconds to milliseconds, no longer than the folding time of small proteins \cite{klepeis2009long,chodera2006long}.
Hence, high-performance computers, efficient computational methods, and also effective models (see below), have been desired and indeed developed to extend the time scale and system size.

\subsection{\label{sec:ff} Force Fields of Molecular Dynamics}

Classical MD assumes that the force on each particle is calculated as a function of particle positions, which is called force field (though often represented in the form of potential energy function).
It is of course approximated function from quantum mechanics \cite{donchev2005quantum}, and model parameters are determined by experiment or quantum chemistry calculation of small systems.
The definition is hence not unique; e.g., bonds between atoms are typically represented by linear springs, however, the stiffness constant varies depending on the force field \cite{weiner1984new}.

For biomolecules such as proteins, nucleic acids, and lipids, there are several force fields in favor such as AMBER \cite{wang2004development,zhang2018evaluation}, CHARMM \cite{hart2012optimization,best2012optimization}, and GROMOS \cite{scott1999gromos}.
Each of these has its strong and weak points.
For example, AMBER provides variants of force fields adapted for different types of systems, popular e.g. for protein folding, and frequently used for nucleic acids. CHARMM is popular e.g. for membrane-bound proteins and has been increasingly used for nucleic acids since the CHARMM36 update with improved accuracy \cite{hart2012optimization}.
An accurate force field specialized for DNA has been recently developed \cite{ivani2016parmbsc1}. These force fields are periodically updated and will be further improved in the future.

\subsection{\label{sec:cg} Coarse-Grained (CG) Models}

As long as the force field is defined, theoretically, we can run MD simulations of any system.
However, the system size and timespan are practically limited by computational costs.
To mitigate the problem, coarse-grained models are often used \cite{noid2013perspective,izvekov2005multiscale,tozzini2005coarse}.
Coarse-graining reduces the number of particles (or variables) in the model,
e.g., atoms in an amino-acid residue in the protein are substituted by a single particle \cite{takada2012coarse,takada2015modeling};
which can decrease the calculation of interactions and also increase the time-step allowed,
drastically saving the computation.
Hence, various coarse-grained models for proteins and nucleic acids have been proposed.
Of course, their accuracy and application are limited due to the simplified form;
e.g., the conformation of the side-chain is lost in the amino-acid level coarse-graining,
and hence microscopic details of interactions cannot be considered \cite{wagner2017extending}.
Currently, the best choice depends on the problem; while all-atom models are adopted for a wide variety of biomolecules, efforts to improve coarse-grained models of proteins and nucleic acids have been taken (including ours \cite{isami2015simple,kameda20171,togashi2018coarse,amyot2019analyzing}).

\subsection{\label{sec:extend} Extended Ensemble Methods}

MD simulation numerically reproduces time series of coordinates, i.e. motion of the molecule.
Considering the snapshots as samples of the structural ensemble (and assuming the ergodicity),
we can evaluate statistics or physical quantities,
e.g., frequency of secondary structure formation in proteins \cite{minary2004long,leimkuhler2016molecular}.
MD simulation in equilibrium can however not always explore the structural space enough to estimate properties \cite{hamelryck2006sampling,rodinger2005enhancing}.
In MD simulation, the initial conformation is usually set according to an experimentally known structure.
Suppose that the molecule takes another stable conformation, and these two conformations are separated by a high energetic barrier.
Then the system cannot reach the other conformation within a realistic amount of computation (i.e. ergodicity is not fulfilled), because the transition state in the middle at high free energy only scarcely appears \cite{elber1987multiple,beck2004methods,henin2004overcoming,muller2013basics,minary2004long}.
Such free energy barriers are one of the biggest issues in MD simulation, and countermeasures have been invented \cite{hamelberg2004accelerated}.

The term extended-ensemble (or generalized-ensemble) method was coined, for those methods using artificial ensemble intended to tackle this problem, typically combined with Monte Carlo and also MD simulation.
Here we briefly introduce the idea without technical details.
These methods make it easier to overcome such barriers by artificially making high-energy states appear more frequently than in the canonical ensemble.
Multicanonical method and simulated tempering method are traditional ones among them.
The replica-exchange method \cite{sugita1999replica,laghaei2011structure,zhou2007replica} and umbrella sampling method \cite{kastner2011umbrella,beutler1994computation,harvey1987umbrella} are popularly used, and a number of their variants have been introduced.
The replica-exchange method is a parallel version of simulated tempering (thus also known as parallel tempering),
in which multiple replicas of the same system are simulated at different temperatures in parallel
(now also called temperature replica-exchange, as there are variants changing a parameter other than temperature).
Between two temperatures, the replicas are exchanged at a probability determined not to violate the detailed balance, to attain the equilibrium.
By using an appropriate set of temperatures, the system can escape from (meta)stable states when at a high temperature.
Replica-exchange MD simulation is now routinely used for various problems, e.g. protein folding \cite{rhee2003multiplexed,nymeyer2004atomic}.
The umbrella sampling method is another way, which cancels the barrier by additional potential energy.
The system behavior is described by generalized coordinates (e.g. end-to-end or domain-to-domain distance), and a bias potential function depending on these coordinates is artificially introduced \cite{best2005reaction,ma2005automatic}.
Then, conformational states in a certain range of the coordinates can be intensively explored.
By using multiple coordinate ranges, states including the barrier region can be broadly sampled;
the probability distribution (and hence the free energy landscape) can be recalculated by canceling the effect of the bias potential \cite{souaille2001extension,kumar1992weighted}.
Although these methods are frequently used for proteins, application to systems including nucleic acids has some intrinsic problems (discussed in the last section).

\section{\label{sec:dna} Molecular Dynamics Analysis of Nucleic Acids}

In this section, we focus on MD studies on simplex nucleic acids such as DNA and RNA.
MD simulation starts from a certain initial structure, which is often taken from an experimentally known structure.
In the case of proteins, as protein folding is generally a slow process, MD simulation usually starts from an already folded structure, typically using structural data deposited in the protein data bank (PDB); homology modeling is applied when necessary.
In contrast, only limited cases of simplex nucleic acid structures are experimentally determined (some short segments are registered in PDB), as they, in general, do not stably fold into a specific structure.
As for the double-stranded structure of nucleic acids e.g. B-DNA, alternatively, the structure can be modeled by stacking base-pairs one by one.
More specifically, the configuration of the base-pair (i.e. translation and rotation from the previous one) is determined by base-step parameters depending on the nucleotides; e.g. X3DNA \cite{lu20033dna,lu20083dna}.
Double-stranded DNA structure for theoretically any nucleotide sequences can be constructed in this way,
and indeed widely used \cite{freeman2014coarse,kameda20171}.
These experimental and predicted structures are applicable for MD simulation of nucleic acids in many cases.

Besides common B-DNA, other structures such as A- and Z-DNA are observed under special conditions \cite{dickerson19925,bukowiecka2006dna}, and their formation or transition from B-DNA has been studied \cite{vargason2001crystallographic,banavali2005free}.
Such structural transitions are also a target of MD simulation;
particularly their solution condition and free energy barrier have been discussed \cite{banavali2005free,kastenholz2006transition,cheatham1997insight}.
Excessive mechanical stress is suggested to induce transitions between these DNA structures \cite{kannan2006b,cerny2008double,knee2008spectroscopic}.
Transitions depending on the chemical modification of DNA have been also studied by MD simulation \cite{temiz2012role}.

Chemical modification frequently occurs in eukaryotic genome DNA and acts as a biochemical signal \cite{jones2001role,rausch2020dna}. Methylation of DNA is broadly observed and biologically important; methylation at C$^5$ of cytosine in particular.
5-methylated cytosine (5mC or mC) in CpG islands (i.e. regions where CpG dinucleotides are condensed) is well known as gene silencing signal, regulating gene expression crucial e.g. for differentiation \cite{baylin2005dna,curradi2002molecular,suzuki2008dna}.
From a physical point of view, effects of methylation have been investigated also by MD simulation;
ranging from microscopic interactions with nearby atoms and hydration \cite{liebl2019methyl,kameda2021structural,teng2018effect,furukawa2021structural} to functional behavior such as strand separation \cite{severin2013effects} and homologous sequence recognition \cite{yoo2016direct}; also relevant to nucleosome formation and positioning \cite{ngo2016effects,portella2013understanding}.
Mechanical aspects of DNA may play significant roles in cellular processes and will be further studied by MD simulation.

Compared with DNA, little is known about the physical properties of RNA, either by experiment or simulation.
Simplex RNA typically does not form a specific structure like DNA double-strands.
The backbone of RNA is flexible, with some complementary sequences forming base-pairs;
its folding pattern is not simply related to the nucleotide sequence,
and hence methods to determine the structure have been sought.
Secondary structures of RNA have been the main focus, for which prediction methods have been suggested and discussed \cite{waterman1978rna,aviran2011modeling,hamada2014fighting}.
Structures of RNA complexed with proteins have been determined by crystallography and their folds have been analyzed \cite{lu2015dssr,hanson2017dssr,lu2020dssr}.
Although prediction of RNA structure is still under development, three-dimensional structure of RNA is investigated e.g. in the context of riboswitches \cite{breaker2011prospects,antunes2018using,domin2017applicability}.
Recently, non-coding RNA has been intensively studied and its function e.g. in gene regulation has been elucidated.
In these fields, the roles of RNA as molecular machines will be focused on, for which the structure-function relationship is important.
Hence, as well as proteins, methods to predict RNA structure from the nucleotide sequence are desired.

MD simulation is also applied to RNA molecules with experimentally known structures.
Still, there is room for improvement of e.g. force field accuracy \cite{vsponer2017understand} and analysis of dynamics \cite{herschlag2018story}.
Experimental data for calibration (e.g. crystallographic B-factor) is still not satisfactory and more data sets would be valuable for the refinement of force fields \cite{buck2006importance,krieger2004making}.
In the analysis, to associate the dynamics and function of the molecule,
the definition of modes of motion, or reaction coordinates, is important,
however lacks consensus for RNA;
in contrast to proteins where functional domains are known in many cases.
In other words, single-stranded RNA shows diverse behavior, calling for novel experiments and theories.

\section{\label{sec:nuc} Molecular Dynamics Analysis of Nucleosomes}

Nucleic acids do not always work alone, and sometimes form a complex with proteins.
For example, chromatin is composed of nucleosomes, i.e. complex of DNA and histone proteins.
In this section, we overview recent MD studies on nucleosomes.
Nucleosome structures have been determined by crystallography and cryoEM;
now many data sets are found in PDB, for different histone variants and nucleotide sequences or with binding proteins.
Conformational determination of multiple nucleosomes along the DNA is enabled by cryoEM.
How to utilize these data is a current issue for theoretical and computational biologists.

Nevertheless, studies on sequence dependence are seldom,
as these experiments employ special sequences in most cases, required for stable nucleosome formation.
Recently, nucleosome sliding has been discussed using MD simulation \cite{brandani2018dna,lequieu2017silico,niina2017sequence}, showing that dynamics of the DNA double-strand depends on the nucleotide sequence, which thus alters the sliding behavior.
In these studies, coarse-grained models of DNA and protein are used.
The use of all-atom models to include more details e.g. chemical modification would be interesting, however, detection of subtle differences may require huge computational costs, partly due to the high stability of nucleosome structure.

Unwrapping of nucleosomes has been also investigated, either by all-atom or coarse-grained models,
as it is important for regulation of the nucleosome structure and accessibility to nucleosomal DNA;
particularly interesting in the context of pioneer factors interacting with nucleosomal DNA and affecting transcription \cite{kagawa2021structural}.
Energetic barriers (i.e. resistance to unwrapping) and roles of histones have been surveyed \cite{armeev2021histone,kono2018free,kono2019free,kono2020nucleosome}.
The stable barrel-like structure makes it easier to characterize nucleosome motion, e.g. using the end-to-end distance of DNA as a coordinate, in contrast to the aforementioned irregular RNA structure; the dynamics is often analyzed based on those structural coordinates.
The behavior of neighboring or overlapping nucleosomes (structure determined by cryoEM) is also a target of MD analysis in detail \cite{matsumoto2020structural}.

Systems of single to several nucleosomes are now good targets of MD simulations.
Currently, the majority of MD studies of nucleosomes focus on histones rather than DNA.
For example, effects of histone modification (e.g. methylation at specific positions of the tail) on the nucleosome dynamics have been investigated from the physical point of view \cite{gatchalian2017accessibility,ikebe2016h3}, and are largely consistent with biochemical studies.
Roles of histone tails in e.g. unwrapping \cite{armeev2021histone} or alignment \cite{ishida2017h4} of nucleosomes have been elucidated.
The dynamics of the tails affects the overall structural stability of nucleosomes \cite{erler2014role,li2016distinct,bignon2021dynamic,medina2021unraveling,bendandi2020role,huertas2021histone}; also in incomplete nucleosome structure lacking some histones \cite{kameda2019histone}.

\section{\label{sec:sum} Summary and Future Perspectives}

In this review, we introduced analysis methods and examples for the structural dynamics of molecular systems including nucleic acids.
Overall, we focused on efficient sampling methods for the conformational states,
due to the prohibitive computational cost of all-atom MD simulation over physiological timescales.
As nucleic acids are becoming targets of bioengineering and drug discovery (e.g. mRNA vaccines),
dynamics of nucleic acids cannot be ignored, and effective computational methods for that will be desired.

Many proteins, although not all, form a specific fold structure,
which could be characterized by e.g. secondary structures or functional domains.
In contrast, nucleic acids do not usually fold into a single structure;
hence it is nontrivial how to define structural states or order parameters.
The double-stranded DNA structure is typically characterized by local (base-pair or base-step) structural variables,
which cannot represent meso- to macro-scopic properties.
Folding of RNA is even more complicated, without a consensus of methods for structural analysis.
Recent advances in the 3D structure prediction of proteins,
represented by AlphaFold2 \cite{tunyasuvunakool2021highly,jumper2021highly} and RoseTTAFold,
have attracted attention in the field of structural biology.
Machine learning approaches used in these works owe much to huge structural data stored in the Protein Data Bank.
Prediction of structure and dynamics of nucleic acids will demand more accumulation of data in the future.

Double-stranded DNA usually denatures at 350--360 K.
Once separated, these strands do not revert to the original double-strand structure in a typical timespan of MD simulation.
Therefore, sampling methods involving complete separation of strands, e.g. temperature replica-exchange MD, cannot work efficiently.
To mitigate this problem, we have recently proposed a method to estimate the free energy landscape
for the binding of base-pairs, without destroying the structure, by applying the adaptive biasing force MD \cite{kameda2021free},
although the system size is limited by its relatively large computational cost.
Using this method, we demonstrated the binding dynamics of codon and anticodon in the ribosome (pre-initiation complex),
to show the mechanical basis of start-codon recognition;
the estimated affinity is consistent with the initiation frequency observed by experiments.
This method would be useful for the analysis of nucleotide-dependent behavior in other systems.

Biopolymers take an innumerable variety, generated by sequences of e.g. nucleotides or amino acids;
which means, only a very limited portion of possible cases can be studied.
This is especially problematic when the molecular function or structure-function relationship is unclear.
Towards universality in biology, how can we effectually survey these molecules?
One of the keys will be informatics, which is strong for comparing items represented by sequences,
although physicochemical properties are often omitted.
The combination of computational biophysics and bioinformatics (including ``multi-omics'' analysis)
will be a powerful tool for the analysis of biomolecular dynamics.
We wish this field will be cooperatively cultivated in the future.

\section*{Author Contributions}
{TK, AA, and YT wrote the paper.}

\section*{Acknowledgments}
{The authors are grateful to S. Tate, T. Yamamoto, N. Sakamoto, M. M. Suzuki, I. Nikaido, R. Erban, K. Asano, S. Shinkai, and H. Nishimori for fruitful discussions. YT's work was supported by JSPS KAKENHI Grant Number JP18KK0388.}

\section*{Conflicts of Interest}
{The authors declare that the research was conducted in the absence of any commercial or financial relationships that could be construed as a potential conflict of interest.}


\begin{thebibliography}{100}

\bibitem{boutet2012high}
S{\'e}bastien Boutet, Lukas Lomb, Garth~J Williams, Thomas~RM Barends, Andrew
  Aquila, R~Bruce Doak, Uwe Weierstall, Daniel~P DePonte, Jan Steinbrener,
  Robert~L Shoeman, et~al.
\newblock High-resolution protein structure determination by serial femtosecond
  crystallography.
\newblock {\em Science}, 337(6092):362--364, 2012.

\bibitem{spence2012x}
JCH Spence, Uwe Weierstall, and HN~Chapman.
\newblock X-ray lasers for structural and dynamic biology.
\newblock {\em Reports on Progress in Physics}, 75(10):102601, 2012.

\bibitem{martin2016serial}
Jose~M Martin-Garcia, Chelsie~E Conrad, Jesse Coe, Shatabdi Roy-Chowdhury, and
  Petra Fromme.
\newblock Serial femtosecond crystallography: A revolution in structural
  biology.
\newblock {\em Archives of biochemistry and biophysics}, 602:32--47, 2016.

\bibitem{kay2011nmr}
Lewis~E Kay.
\newblock Nmr studies of protein structure and dynamics.
\newblock {\em Journal of Magnetic Resonance}, 213(2):477--491, 2011.

\bibitem{prestegard1998new}
JH~Prestegard.
\newblock New techniques in structural {NMR}---anisotropic interactions.
\newblock {\em Nature structural biology}, 5(7):517--522, 1998.

\bibitem{topf2008protein}
Maya Topf, Keren Lasker, Ben Webb, Haim Wolfson, Wah Chiu, and Andrej Sali.
\newblock Protein structure fitting and refinement guided by cryo-em density.
\newblock {\em Structure}, 16(2):295--307, 2008.

\bibitem{yip2020atomic}
Ka~Man Yip, Niels Fischer, Elham Paknia, Ashwin Chari, and Holger Stark.
\newblock Atomic-resolution protein structure determination by cryo-em.
\newblock {\em Nature}, 587(7832):157--161, 2020.

\bibitem{shi2013three}
Dan Shi, Brent~L Nannenga, Matthew~G Iadanza, and Tamir Gonen.
\newblock Three-dimensional electron crystallography of protein microcrystals.
\newblock {\em Elife}, 2:e01345, 2013.

\bibitem{nogales2016development}
Eva Nogales.
\newblock The development of cryo-em into a mainstream structural biology
  technique.
\newblock {\em Nature methods}, 13(1):24--27, 2016.

\bibitem{ho2018rapid}
Phuong~T Ho and Vijay~S Reddy.
\newblock Rapid increase of near atomic resolution virus capsid structures
  determined by cryo-electron microscopy.
\newblock {\em Journal of structural biology}, 201(1):1--4, 2018.

\bibitem{raunser2017cryo}
Stefan Raunser.
\newblock Cryo-em revolutionizes the structure determination of biomolecules.
\newblock {\em Angewandte Chemie International Edition}, 56(52):16450--16452,
  2017.

\bibitem{branden2012introduction}
Carl~Ivar Branden and John Tooze.
\newblock {\em Introduction to protein structure}.
\newblock Garland Science, 2012.

\bibitem{sadowski2009sequence}
MI~Sadowski and DT~Jones.
\newblock The sequence--structure relationship and protein function prediction.
\newblock {\em Current opinion in structural biology}, 19(3):357--362, 2009.

\bibitem{crick1953molecular}
FHC Crick and JD~Watson.
\newblock Molecular structure of deoxypentose nucleic acids.
\newblock {\em Nature}, 171:738--740, 1953.

\bibitem{sim2012modeling}
Adelene~YL Sim, Peter Minary, and Michael Levitt.
\newblock Modeling nucleic acids.
\newblock {\em Current opinion in structural biology}, 22(3):273--278, 2012.

\bibitem{blackburn2006nucleic}
G~Michael Blackburn, Michael~J Gait, David Loakes, David~M Williams, Jane~A
  Grasby, Martin Egli, Andy Flavell, Stephanie Allen, Julie Fisher, Anna~Marie
  Pyle, et~al.
\newblock {\em Nucleic acids in chemistry and biology}.
\newblock Royal Society of Chemistry, 2006.

\bibitem{mccammon1988dynamics}
J~Andrew McCammon and Stephen~C Harvey.
\newblock {\em Dynamics of proteins and nucleic acids}.
\newblock Cambridge University Press, 1988.

\bibitem{rief1999sequence}
Matthias Rief, Hauke Clausen-Schaumann, and Hermann~E Gaub.
\newblock Sequence-dependent mechanics of single dna molecules.
\newblock {\em Nature structural biology}, 6(4):346--349, 1999.

\bibitem{ma2016anisotropy}
Ning Ma and Arjan van~der Vaart.
\newblock Anisotropy of b-dna groove bending.
\newblock {\em Journal of the American Chemical Society}, 138(31):9951--9958,
  2016.

\bibitem{davey2002solvent}
Curt~A Davey, David~F Sargent, Karolin Luger, Armin~W Maeder, and Timothy~J
  Richmond.
\newblock Solvent mediated interactions in the structure of the nucleosome core
  particle at 1.9 {\aa} resolution.
\newblock {\em Journal of molecular biology}, 319(5):1097--1113, 2002.

\bibitem{vasudevan2010crystal}
Dileep Vasudevan, Eugene~YD Chua, and Curt~A Davey.
\newblock Crystal structures of nucleosome core particles containing the `601'
  strong positioning sequence.
\newblock {\em Journal of molecular biology}, 403(1):1--10, 2010.

\bibitem{tan2011nucleosome}
Song Tan and Curt~A Davey.
\newblock Nucleosome structural studies.
\newblock {\em Current opinion in structural biology}, 21(1):128--136, 2011.

\bibitem{luger2005nucleosome}
Karolin Luger and Jeffrey~C Hansen.
\newblock Nucleosome and chromatin fiber dynamics.
\newblock {\em Current opinion in structural biology}, 15(2):188--196, 2005.

\bibitem{dai2020evolving}
Ziwei Dai, Vijyendra Ramesh, and Jason~W Locasale.
\newblock The evolving metabolic landscape of chromatin biology and
  epigenetics.
\newblock {\em Nature Reviews Genetics}, 21(12):737--753, 2020.

\bibitem{hihara2012local}
Saera Hihara, Chan-Gi Pack, Kazunari Kaizu, Tomomi Tani, Tomo Hanafusa, Tadasu
  Nozaki, Satoko Takemoto, Tomohiko Yoshimi, Hideo Yokota, Naoko Imamoto,
  et~al.
\newblock Local nucleosome dynamics facilitate chromatin accessibility in
  living mammalian cells.
\newblock {\em Cell reports}, 2(6):1645--1656, 2012.

\bibitem{rao2010protein}
Francesco Rao and Martin Karplus.
\newblock Protein dynamics investigated by inherent structure analysis.
\newblock {\em Proceedings of the National Academy of Sciences},
  107(20):9152--9157, 2010.

\bibitem{frenkel2001understanding}
Daan Frenkel and Berend Smit.
\newblock {\em Understanding molecular simulation: from algorithms to
  applications}, volume~1.
\newblock Elsevier, 2001.

\bibitem{kofke2004getting}
David~A Kofke.
\newblock Getting the most from molecular simulation.
\newblock {\em Molecular Physics}, 102(4):405--420, 2004.

\bibitem{bonomi2019biomolecular}
Massimiliano Bonomi and Carlo Camilloni.
\newblock {\em Biomolecular Simulations}.
\newblock Springer, 2019.

\bibitem{huggins2019biomolecular}
David~J Huggins, Philip~C Biggin, Marc~A D{\"a}mgen, Jonathan~W Essex, Sarah~A
  Harris, Richard~H Henchman, Syma Khalid, Antonija Kuzmanic, Charles~A
  Laughton, Julien Michel, et~al.
\newblock Biomolecular simulations: From dynamics and mechanisms to
  computational assays of biological activity.
\newblock {\em Wiley Interdisciplinary Reviews: Computational Molecular
  Science}, 9(3):e1393, 2019.

\bibitem{karplus2002molecular}
Martin Karplus and J~Andrew McCammon.
\newblock Molecular dynamics simulations of biomolecules.
\newblock {\em Nature structural biology}, 9(9):646--652, 2002.

\bibitem{karplus1990molecular}
Martin Karplus and Gregory~A Petsko.
\newblock Molecular dynamics simulations in biology.
\newblock {\em Nature}, 347(6294):631--639, 1990.

\bibitem{bao2002mechanics}
Gang Bao.
\newblock Mechanics of biomolecules.
\newblock {\em Journal of the Mechanics and Physics of Solids},
  50(11):2237--2274, 2002.

\bibitem{gogonea2001new}
Valentin Gogonea, Dimas Su{\'a}rez, Arjan van~der Vaart, and Kenneth~M Merz~Jr.
\newblock New developments in applying quantum mechanics to proteins.
\newblock {\em Current opinion in structural biology}, 11(2):217--223, 2001.

\bibitem{stock2009classical}
Gerhard Stock.
\newblock Classical simulation of quantum energy flow in biomolecules.
\newblock {\em Physical review letters}, 102(11):118301, 2009.

\bibitem{donchev2005quantum}
AG~Donchev, VD~Ozrin, MV~Subbotin, OV~Tarasov, and VI~Tarasov.
\newblock A quantum mechanical polarizable force field for biomolecular
  interactions.
\newblock {\em Proceedings of the National Academy of Sciences},
  102(22):7829--7834, 2005.

\bibitem{dunning1977gaussian}
Thom~H Dunning and P~Jeffrey Hay.
\newblock Gaussian basis sets for molecular calculations.
\newblock In {\em Methods of electronic structure theory}, pages 1--27.
  Springer, 1977.

\bibitem{hansson2002molecular}
Tomas Hansson, Chris Oostenbrink, and WilfredF van Gunsteren.
\newblock Molecular dynamics simulations.
\newblock {\em Current opinion in structural biology}, 12(2):190--196, 2002.

\bibitem{karplus2005molecular}
Martin Karplus and John Kuriyan.
\newblock Molecular dynamics and protein function.
\newblock {\em Proceedings of the National Academy of Sciences},
  102(19):6679--6685, 2005.

\bibitem{binder2004molecular}
Kurt Binder, J{\"u}rgen Horbach, Walter Kob, Wolfgang Paul, and Fathollah
  Varnik.
\newblock Molecular dynamics simulations.
\newblock {\em Journal of Physics: Condensed Matter}, 16(5):S429, 2004.

\bibitem{li2005basic}
Ju~Li.
\newblock Basic molecular dynamics.
\newblock In {\em Handbook of Materials Modeling}, pages 565--588. Springer,
  2005.

\bibitem{paterlini1998constant}
M~Germana Paterlini and David~M Ferguson.
\newblock Constant temperature simulations using the langevin equation with
  velocity verlet integration.
\newblock {\em Chemical Physics}, 236(1-3):243--252, 1998.

\bibitem{tuckerman1992reversible}
MBBJM Tuckerman, Bruce~J Berne, and Glenn~J Martyna.
\newblock Reversible multiple time scale molecular dynamics.
\newblock {\em The Journal of chemical physics}, 97(3):1990--2001, 1992.

\bibitem{klepeis2009long}
John~L Klepeis, Kresten Lindorff-Larsen, Ron~O Dror, and David~E Shaw.
\newblock Long-timescale molecular dynamics simulations of protein structure
  and function.
\newblock {\em Current opinion in structural biology}, 19(2):120--127, 2009.

\bibitem{chodera2006long}
John~D Chodera, William~C Swope, Jed~W Pitera, and Ken~A Dill.
\newblock Long-time protein folding dynamics from short-time molecular dynamics
  simulations.
\newblock {\em Multiscale Modeling \& Simulation}, 5(4):1214--1226, 2006.

\bibitem{weiner1984new}
Scott~J Weiner, Peter~A Kollman, David~A Case, U~Chandra Singh, Caterina Ghio,
  Guliano Alagona, Salvatore Profeta, and Paul Weiner.
\newblock A new force field for molecular mechanical simulation of nucleic
  acids and proteins.
\newblock {\em Journal of the American Chemical Society}, 106(3):765--784,
  1984.

\bibitem{wang2004development}
Junmei Wang, Romain~M Wolf, James~W Caldwell, Peter~A Kollman, and David~A
  Case.
\newblock Development and testing of a general amber force field.
\newblock {\em Journal of computational chemistry}, 25(9):1157--1174, 2004.

\bibitem{zhang2018evaluation}
Yushan Zhang, Yong Zhang, Mark~J McCready, and Edward~J Maginn.
\newblock Evaluation and refinement of the general amber force field for
  nineteen pure organic electrolyte solvents.
\newblock {\em Journal of Chemical \& Engineering Data}, 63(9):3488--3502,
  2018.

\bibitem{hart2012optimization}
Katarina Hart, Nicolas Foloppe, Christopher~M Baker, Elizabeth~J Denning,
  Lennart Nilsson, and Alexander~D MacKerell~Jr.
\newblock Optimization of the charmm additive force field for dna: Improved
  treatment of the bi/bii conformational equilibrium.
\newblock {\em Journal of chemical theory and computation}, 8(1):348--362,
  2012.

\bibitem{best2012optimization}
Robert~B Best, Xiao Zhu, Jihyun Shim, Pedro~EM Lopes, Jeetain Mittal, Michael
  Feig, and Alexander~D MacKerell~Jr.
\newblock Optimization of the additive charmm all-atom protein force field
  targeting improved sampling of the backbone $\phi$, $\psi$ and side-chain
  $\chi$1 and $\chi$2 dihedral angles.
\newblock {\em Journal of chemical theory and computation}, 8(9):3257--3273,
  2012.

\bibitem{scott1999gromos}
Walter~RP Scott, Philippe~H H{\"u}nenberger, Ilario~G Tironi, Alan~E Mark,
  Salomon~R Billeter, Jens Fennen, Andrew~E Torda, Thomas Huber, Peter
  Kr{\"u}ger, and Wilfred~F van Gunsteren.
\newblock The gromos biomolecular simulation program package.
\newblock {\em The Journal of Physical Chemistry A}, 103(19):3596--3607, 1999.

\bibitem{ivani2016parmbsc1}
Ivan Ivani, Pablo~D Dans, Agnes Noy, Alberto P{\'e}rez, Ignacio Faustino, Adam
  Hospital, J{\"u}rgen Walther, Pau Andrio, Ramon Go{\~n}i, Alexandra
  Balaceanu, et~al.
\newblock Parmbsc1: a refined force field for dna simulations.
\newblock {\em Nature methods}, 13(1):55--58, 2016.

\bibitem{noid2013perspective}
William~George Noid.
\newblock Perspective: Coarse-grained models for biomolecular systems.
\newblock {\em The Journal of chemical physics}, 139(9):09B201\_1, 2013.

\bibitem{izvekov2005multiscale}
Sergei Izvekov and Gregory~A Voth.
\newblock A multiscale coarse-graining method for biomolecular systems.
\newblock {\em The Journal of Physical Chemistry B}, 109(7):2469--2473, 2005.

\bibitem{tozzini2005coarse}
Valentina Tozzini.
\newblock Coarse-grained models for proteins.
\newblock {\em Current opinion in structural biology}, 15(2):144--150, 2005.

\bibitem{takada2012coarse}
Shoji Takada.
\newblock Coarse-grained molecular simulations of large biomolecules.
\newblock {\em Current opinion in structural biology}, 22(2):130--137, 2012.

\bibitem{takada2015modeling}
Shoji Takada, Ryo Kanada, Cheng Tan, Tsuyoshi Terakawa, Wenfei Li, and Hiroo
  Kenzaki.
\newblock Modeling structural dynamics of biomolecular complexes by
  coarse-grained molecular simulations.
\newblock {\em Accounts of chemical research}, 48(12):3026--3035, 2015.

\bibitem{wagner2017extending}
Jacob~W Wagner, Thomas Dannenhoffer-Lafage, Jaehyeok Jin, and Gregory~A Voth.
\newblock Extending the range and physical accuracy of coarse-grained models:
  Order parameter dependent interactions.
\newblock {\em The Journal of chemical physics}, 147(4):044113, 2017.

\bibitem{isami2015simple}
Shuhei Isami, Naoaki Sakamoto, Hiraku Nishimori, and Akinori Awazu.
\newblock Simple elastic network models for exhaustive analysis of long
  double-stranded dna dynamics with sequence geometry dependence.
\newblock {\em PLoS one}, 10(12):e0143760, 2015.

\bibitem{kameda20171}
Takeru Kameda, Shuhei Isami, Yuichi Togashi, Hiraku Nishimori, Naoaki Sakamoto,
  and Akinori Awazu.
\newblock The 1-particle-per-k-nucleotides (1pkn) elastic network model of dna
  dynamics with sequence-dependent geometry.
\newblock {\em Frontiers in physiology}, 8:103, 2017.

\bibitem{togashi2018coarse}
Yuichi Togashi and Holger Flechsig.
\newblock Coarse-grained protein dynamics studies using elastic network models.
\newblock {\em International journal of molecular sciences}, 19(12):3899, 2018.

\bibitem{amyot2019analyzing}
Romain Amyot, Yuichi Togashi, and Holger Flechsig.
\newblock Analyzing fluctuation properties in protein elastic networks with
  sequence-specific and distance-dependent interactions.
\newblock {\em Biomolecules}, 9(10):549, 2019.

\bibitem{minary2004long}
P~Minary, ME~Tuckerman, and GJ~Martyna.
\newblock Long time molecular dynamics for enhanced conformational sampling in
  biomolecular systems.
\newblock {\em Physical review letters}, 93(15):150201, 2004.

\bibitem{leimkuhler2016molecular}
Ben Leimkuhler and Charles Matthews.
\newblock {\em Molecular Dynamics.}
\newblock Springer, 2016.

\bibitem{hamelryck2006sampling}
Thomas Hamelryck, John~T Kent, and Anders Krogh.
\newblock Sampling realistic protein conformations using local structural bias.
\newblock {\em PLoS Comput Biol}, 2(9):e131, 2006.

\bibitem{rodinger2005enhancing}
Tomas Rodinger and R{\'e}gis Pom{\`e}s.
\newblock Enhancing the accuracy, the efficiency and the scope of free energy
  simulations.
\newblock {\em Current opinion in structural biology}, 15(2):164--170, 2005.

\bibitem{elber1987multiple}
R~Elber and Martin Karplus.
\newblock Multiple conformational states of proteins: a molecular dynamics
  analysis of myoglobin.
\newblock {\em Science}, 235(4786):318--321, 1987.

\bibitem{beck2004methods}
David~AC Beck and Valerie Daggett.
\newblock Methods for molecular dynamics simulations of protein
  folding/unfolding in solution.
\newblock {\em Methods}, 34(1):112--120, 2004.

\bibitem{henin2004overcoming}
J{\'e}r{\^o}me H{\'e}nin and Christophe Chipot.
\newblock Overcoming free energy barriers using unconstrained molecular
  dynamics simulations.
\newblock {\em The Journal of chemical physics}, 121(7):2904--2914, 2004.

\bibitem{muller2013basics}
Harald~JW Muller-Kirsten.
\newblock {\em Basics of statistical physics}.
\newblock World Scientific Publishing Company, 2013.

\bibitem{hamelberg2004accelerated}
Donald Hamelberg, John Mongan, and J~Andrew McCammon.
\newblock Accelerated molecular dynamics: a promising and efficient simulation
  method for biomolecules.
\newblock {\em The Journal of chemical physics}, 120(24):11919--11929, 2004.

\bibitem{sugita1999replica}
Yuji Sugita and Yuko Okamoto.
\newblock Replica-exchange molecular dynamics method for protein folding.
\newblock {\em Chemical physics letters}, 314(1-2):141--151, 1999.

\bibitem{laghaei2011structure}
Rozita Laghaei, Normand Mousseau, and Guanghong Wei.
\newblock Structure and thermodynamics of amylin dimer studied by
  hamiltonian-temperature replica exchange molecular dynamics simulations.
\newblock {\em The Journal of Physical Chemistry B}, 115(12):3146--3154, 2011.

\bibitem{zhou2007replica}
Ruhong Zhou.
\newblock Replica exchange molecular dynamics method for protein folding
  simulation.
\newblock In {\em Protein Folding Protocols}, pages 205--223. Springer, 2007.

\bibitem{kastner2011umbrella}
Johannes K{\"a}stner.
\newblock Umbrella sampling.
\newblock {\em Wiley Interdisciplinary Reviews: Computational Molecular
  Science}, 1(6):932--942, 2011.

\bibitem{beutler1994computation}
Thomas~C Beutler and Wilfred~F van Gunsteren.
\newblock The computation of a potential of mean force: choice of the biasing
  potential in the umbrella sampling technique.
\newblock {\em The Journal of chemical physics}, 100(2):1492--1497, 1994.

\bibitem{harvey1987umbrella}
Stephen~C Harvey and M~Prabhakaran.
\newblock Umbrella sampling: avoiding possible artifacts and statistical
  biases.
\newblock {\em Journal of Physical Chemistry}, 91(18):4799--4801, 1987.

\bibitem{rhee2003multiplexed}
Young~Min Rhee and Vijay~S Pande.
\newblock Multiplexed-replica exchange molecular dynamics method for protein
  folding simulation.
\newblock {\em Biophysical journal}, 84(2):775--786, 2003.

\bibitem{nymeyer2004atomic}
Hugh Nymeyer, S~Gnanakaran, and Angel~E Garcia.
\newblock Atomic simulations of protein folding, using the replica exchange
  algorithm.
\newblock {\em Methods in enzymology}, 383:119--149, 2004.

\bibitem{best2005reaction}
Robert~B Best and Gerhard Hummer.
\newblock Reaction coordinates and rates from transition paths.
\newblock {\em Proceedings of the National Academy of Sciences},
  102(19):6732--6737, 2005.

\bibitem{ma2005automatic}
Ao~Ma and Aaron~R Dinner.
\newblock Automatic method for identifying reaction coordinates in complex
  systems.
\newblock {\em The Journal of Physical Chemistry B}, 109(14):6769--6779, 2005.

\bibitem{souaille2001extension}
Marc Souaille and Beno{\i}t Roux.
\newblock Extension to the weighted histogram analysis method: combining
  umbrella sampling with free energy calculations.
\newblock {\em Computer physics communications}, 135(1):40--57, 2001.

\bibitem{kumar1992weighted}
Shankar Kumar, John~M Rosenberg, Djamal Bouzida, Robert~H Swendsen, and Peter~A
  Kollman.
\newblock The weighted histogram analysis method for free-energy calculations
  on biomolecules. i. the method.
\newblock {\em Journal of computational chemistry}, 13(8):1011--1021, 1992.

\bibitem{lu20033dna}
Xiang-Jun Lu and Wilma~K Olson.
\newblock 3dna: a software package for the analysis, rebuilding and
  visualization of three-dimensional nucleic acid structures.
\newblock {\em Nucleic acids research}, 31(17):5108--5121, 2003.

\bibitem{lu20083dna}
Xiang-Jun Lu and Wilma~K Olson.
\newblock 3dna: a versatile, integrated software system for the analysis,
  rebuilding and visualization of three-dimensional nucleic-acid structures.
\newblock {\em Nature protocols}, 3(7):1213, 2008.

\bibitem{freeman2014coarse}
Gordon~S Freeman, Daniel~M Hinckley, Joshua~P Lequieu, Jonathan~K Whitmer, and
  Juan~J de~Pablo.
\newblock Coarse-grained modeling of dna curvature.
\newblock {\em The Journal of chemical physics}, 141(16):10B615\_1, 2014.

\bibitem{dickerson19925}
Richard~E Dickerson.
\newblock [5] dna structure from a to z.
\newblock {\em Methods in enzymology}, 211:67--111, 1992.

\bibitem{bukowiecka2006dna}
Ma{\l}gorzata Bukowiecka-Matusiak and Lucyna~A Wo{\'z}niak.
\newblock Dna structure from a to z--biological implications of structural
  diversity of dna.
\newblock {\em Postepy biochemii}, 52(3):229--238, 2006.

\bibitem{vargason2001crystallographic}
Jeffrey~M Vargason, Keith Henderson, and P~Shing Ho.
\newblock A crystallographic map of the transition from b-dna to a-dna.
\newblock {\em Proceedings of the National Academy of Sciences},
  98(13):7265--7270, 2001.

\bibitem{banavali2005free}
Nilesh~K Banavali and Beno{\^\i}t Roux.
\newblock Free energy landscape of a-dna to b-dna conversion in aqueous
  solution.
\newblock {\em Journal of the American Chemical Society}, 127(18):6866--6876,
  2005.

\bibitem{kastenholz2006transition}
Mika~A Kastenholz, Thomas~U Schwartz, and Philippe~H H{\"u}nenberger.
\newblock The transition between the b and z conformations of dna investigated
  by targeted molecular dynamics simulations with explicit solvation.
\newblock {\em Biophysical journal}, 91(8):2976--2990, 2006.

\bibitem{cheatham1997insight}
Thomas~E Cheatham~III and Peter~A Kollman.
\newblock Insight into the stabilization of a-dna by specific ion association:
  spontaneous b-dna to a-dna transitions observed in molecular dynamics
  simulations of d [acccgcgggt] 2 in the presence of hexaamminecobalt (iii).
\newblock {\em Structure}, 5(10):1297--1311, 1997.

\bibitem{kannan2006b}
Srinivasaraghavan Kannan, Kai Kohlhoff, and Martin Zacharias.
\newblock B-dna under stress: over-and untwisting of dna during molecular
  dynamics simulations.
\newblock {\em Biophysical journal}, 91(8):2956--2965, 2006.

\bibitem{cerny2008double}
Jiri Cerny, Martin Kabel{\'a}c, and Pavel Hobza.
\newblock Double-helical $\rightarrow$ ladder structural transition in the
  b-dna is induced by a loss of dispersion energy.
\newblock {\em Journal of the American Chemical Society}, 130(47):16055--16059,
  2008.

\bibitem{knee2008spectroscopic}
Kelly~M Knee, Surjit~B Dixit, Colin~Echeverr{\'\i}a Aitken, Sergei Ponomarev,
  DL~Beveridge, and Ishita Mukerji.
\newblock Spectroscopic and molecular dynamics evidence for a sequential
  mechanism for the a-to-b transition in dna.
\newblock {\em Biophysical Journal}, 95(1):257--272, 2008.

\bibitem{temiz2012role}
Nuri~A Temiz, Duncan~E Donohue, Albino Bacolla, Brian~T Luke, and Jack~R
  Collins.
\newblock The role of methylation in the intrinsic dynamics of b-and z-dna.
\newblock {\em PloS one}, 7(4):e35558, 2012.

\bibitem{jones2001role}
Peter~A Jones and Daiya Takai.
\newblock The role of dna methylation in mammalian epigenetics.
\newblock {\em Science}, 293(5532):1068--1070, 2001.

\bibitem{rausch2020dna}
Cathia Rausch, Florian~D Hastert, and M~Cristina Cardoso.
\newblock Dna modification readers and writers and their interplay.
\newblock {\em Journal of molecular biology}, 432(6):1731--1746, 2020.

\bibitem{baylin2005dna}
Stephen~B Baylin.
\newblock Dna methylation and gene silencing in cancer.
\newblock {\em Nature clinical practice Oncology}, 2(1):S4--S11, 2005.

\bibitem{curradi2002molecular}
Michela Curradi, Annalisa Izzo, Gianfranco Badaracco, and Nicoletta
  Landsberger.
\newblock Molecular mechanisms of gene silencing mediated by dna methylation.
\newblock {\em Molecular and cellular biology}, 22(9):3157--3173, 2002.

\bibitem{suzuki2008dna}
Miho~M Suzuki and Adrian Bird.
\newblock Dna methylation landscapes: provocative insights from epigenomics.
\newblock {\em Nature reviews genetics}, 9(6):465--476, 2008.

\bibitem{liebl2019methyl}
Korbinian Liebl and Martin Zacharias.
\newblock How methyl--sugar interactions determine dna structure and
  flexibility.
\newblock {\em Nucleic acids research}, 47(3):1132--1140, 2019.

\bibitem{kameda2021structural}
Takeru Kameda, Miho~M Suzuki, Akinori Awazu, and Yuichi Togashi.
\newblock Structural dynamics of dna depending on methylation pattern.
\newblock {\em Physical Review E}, 103(1):012404, 2021.

\bibitem{teng2018effect}
Xiaojing Teng and Wonmuk Hwang.
\newblock Effect of methylation on local mechanics and hydration structure of
  dna.
\newblock {\em Biophysical journal}, 114(8):1791--1803, 2018.

\bibitem{furukawa2021structural}
Ayako Furukawa, Erik Walinda, Kyohei Arita, and Kenji Sugase.
\newblock Structural dynamics of double-stranded dna with epigenome
  modification.
\newblock {\em Nucleic acids research}, 49(2):1152--1162, 2021.

\bibitem{severin2013effects}
Philip~MD Severin, Xueqing Zou, Klaus Schulten, and Hermann~E Gaub.
\newblock Effects of cytosine hydroxymethylation on dna strand separation.
\newblock {\em Biophysical journal}, 104(1):208--215, 2013.

\bibitem{yoo2016direct}
Jejoong Yoo, Hajin Kim, Aleksei Aksimentiev, and Taekjip Ha.
\newblock Direct evidence for sequence-dependent attraction between
  double-stranded dna controlled by methylation.
\newblock {\em Nature communications}, 7(1):1--7, 2016.

\bibitem{ngo2016effects}
Thuy~TM Ngo, Jejoong Yoo, Qing Dai, Qiucen Zhang, Chuan He, Aleksei
  Aksimentiev, and Taekjip Ha.
\newblock Effects of cytosine modifications on dna flexibility and nucleosome
  mechanical stability.
\newblock {\em Nature communications}, 7(1):1--9, 2016.

\bibitem{portella2013understanding}
Guillem Portella, Federica Battistini, and Modesto Orozco.
\newblock Understanding the connection between epigenetic dna methylation and
  nucleosome positioning from computer simulations.
\newblock {\em PLoS computational biology}, 9(11):e1003354, 2013.

\bibitem{waterman1978rna}
Michael~S Waterman and Temple~F Smith.
\newblock Rna secondary structure: A complete mathematical analysis.
\newblock {\em Mathematical Biosciences}, 42(3-4):257--266, 1978.

\bibitem{aviran2011modeling}
Sharon Aviran, Cole Trapnell, Julius~B Lucks, Stefanie~A Mortimer, Shujun Luo,
  Gary~P Schroth, Jennifer~A Doudna, Adam~P Arkin, and Lior Pachter.
\newblock Modeling and automation of sequencing-based characterization of rna
  structure.
\newblock {\em Proceedings of the National Academy of Sciences},
  108(27):11069--11074, 2011.

\bibitem{hamada2014fighting}
Michiaki Hamada.
\newblock Fighting against uncertainty: an essential issue in bioinformatics.
\newblock {\em Briefings in bioinformatics}, 15(5):748--767, 2014.

\bibitem{lu2015dssr}
Xiang-Jun Lu, Harmen~J Bussemaker, and Wilma~K Olson.
\newblock Dssr: an integrated software tool for dissecting the spatial
  structure of rna.
\newblock {\em Nucleic acids research}, 43(21):e142--e142, 2015.

\bibitem{hanson2017dssr}
Robert~M Hanson and Xiang-Jun Lu.
\newblock Dssr-enhanced visualization of nucleic acid structures in jmol.
\newblock {\em Nucleic acids research}, 45(W1):W528--W533, 2017.

\bibitem{lu2020dssr}
Xiang-Jun Lu.
\newblock Dssr-enabled innovative schematics of 3d nucleic acid structures with
  pymol.
\newblock {\em Nucleic acids research}, 48(13):e74--e74, 2020.

\bibitem{breaker2011prospects}
Ronald~R Breaker.
\newblock Prospects for riboswitch discovery and analysis.
\newblock {\em Molecular cell}, 43(6):867--879, 2011.

\bibitem{antunes2018using}
Deborah Antunes, Natasha~AN Jorge, Ernesto~R Caffarena, and Fabio Passetti.
\newblock Using rna sequence and structure for the prediction of riboswitch
  aptamer: a comprehensive review of available software and tools.
\newblock {\em Frontiers in genetics}, 8:231, 2018.

\bibitem{domin2017applicability}
Gesine Domin, Sven Findei{\ss}, Manja Wachsmuth, Sebastian Will, Peter~F
  Stadler, and Mario M{\"o}rl.
\newblock Applicability of a computational design approach for synthetic
  riboswitches.
\newblock {\em Nucleic acids research}, 45(7):4108--4119, 2017.

\bibitem{vsponer2017understand}
Ji{\v{r}}{\'\i} {\v{S}}poner, Miroslav Krepl, Pavel Ban{\'a}{\v{s}}, Petra
  K{\"u}hrov{\'a}, Marie Zgarbov{\'a}, Petr Jure{\v{c}}ka, Marek Havrila, and
  Michal Otyepka.
\newblock How to understand atomistic molecular dynamics simulations of rna and
  protein--rna complexes?
\newblock {\em Wiley Interdisciplinary Reviews: RNA}, 8(3):e1405, 2017.

\bibitem{herschlag2018story}
Daniel Herschlag, Steve Bonilla, and Namita Bisaria.
\newblock The story of rna folding, as told in epochs.
\newblock {\em Cold Spring Harbor perspectives in biology}, 10(10):a032433,
  2018.

\bibitem{buck2006importance}
Matthias Buck, Sabine Bouguet-Bonnet, Richard~W Pastor, and Alexander~D
  MacKerell~Jr.
\newblock Importance of the cmap correction to the charmm22 protein force
  field: dynamics of hen lysozyme.
\newblock {\em Biophysical journal}, 90(4):L36--L38, 2006.

\bibitem{krieger2004making}
Elmar Krieger, Tom Darden, Sander~B Nabuurs, Alexei Finkelstein, and Gert
  Vriend.
\newblock Making optimal use of empirical energy functions: force-field
  parameterization in crystal space.
\newblock {\em Proteins: Structure, Function, and Bioinformatics},
  57(4):678--683, 2004.

\bibitem{brandani2018dna}
Giovanni~B Brandani, Toru Niina, Cheng Tan, and Shoji Takada.
\newblock Dna sliding in nucleosomes via twist defect propagation revealed by
  molecular simulations.
\newblock {\em Nucleic acids research}, 46(6):2788--2801, 2018.

\bibitem{lequieu2017silico}
Joshua Lequieu, David~C Schwartz, and Juan~J de~Pablo.
\newblock In silico evidence for sequence-dependent nucleosome sliding.
\newblock {\em Proceedings of the National Academy of Sciences},
  114(44):E9197--E9205, 2017.

\bibitem{niina2017sequence}
Toru Niina, Giovanni~B Brandani, Cheng Tan, and Shoji Takada.
\newblock Sequence-dependent nucleosome sliding in rotation-coupled and
  uncoupled modes revealed by molecular simulations.
\newblock {\em PLoS computational biology}, 13(12):e1005880, 2017.

\bibitem{kagawa2021structural}
Wataru Kagawa and Hitoshi Kurumizaka.
\newblock Structural basis for dna sequence recognition by pioneer factors in
  nucleosomes.
\newblock {\em Current Opinion in Structural Biology}, 71:59--64, 2021.

\bibitem{armeev2021histone}
Grigoriy~A Armeev, Anastasiia~S Kniazeva, Galina~A Komarova, Mikhail~P
  Kirpichnikov, and Alexey~K Shaytan.
\newblock Histone dynamics mediate dna unwrapping and sliding in nucleosomes.
\newblock {\em Nature communications}, 12(1):1--15, 2021.

\bibitem{kono2018free}
Hidetoshi Kono, Shun Sakuraba, and Hisashi Ishida.
\newblock Free energy profiles for unwrapping the outer superhelical turn of
  nucleosomal dna.
\newblock {\em PLoS computational biology}, 14(3):e1006024, 2018.

\bibitem{kono2019free}
Hidetoshi Kono, Shun Sakuraba, and Hisashi Ishida.
\newblock Free energy profile for unwrapping outer superhelical turn of cenp-a
  nucleosome.
\newblock {\em Biophysics and physicobiology}, 16:337--343, 2019.

\bibitem{kono2020nucleosome}
Hidetoshi Kono and Hisashi Ishida.
\newblock Nucleosome unwrapping and unstacking.
\newblock {\em Current Opinion in Structural Biology}, 64:119--125, 2020.

\bibitem{matsumoto2020structural}
Atsushi Matsumoto, Masaaki Sugiyama, Zhenhai Li, Anne Martel, Lionel Porcar,
  Rintaro Inoue, Daiki Kato, Akihisa Osakabe, Hitoshi Kurumizaka, and Hidetoshi
  Kono.
\newblock Structural studies of overlapping dinucleosomes in solution.
\newblock {\em Biophysical journal}, 118(9):2209--2219, 2020.

\bibitem{gatchalian2017accessibility}
Jovylyn Gatchalian, Xiaodong Wang, Jinzen Ikebe, Khan~L Cox, Adam~H Tencer,
  Yi~Zhang, Nathaniel~L Burge, Luo Di, Matthew~D Gibson, Catherine~A Musselman,
  et~al.
\newblock Accessibility of the histone h3 tail in the nucleosome for binding of
  paired readers.
\newblock {\em Nature communications}, 8(1):1--10, 2017.

\bibitem{ikebe2016h3}
Jinzen Ikebe, Shun Sakuraba, and Hidetoshi Kono.
\newblock H3 histone tail conformation within the nucleosome and the impact of
  k14 acetylation studied using enhanced sampling simulation.
\newblock {\em PLoS computational biology}, 12(3):e1004788, 2016.

\bibitem{ishida2017h4}
Hisashi Ishida and Hidetoshi Kono.
\newblock H4 tails potentially produce the diversity in the orientation of two
  nucleosomes.
\newblock {\em Biophysical journal}, 113(5):978--990, 2017.

\bibitem{erler2014role}
Jochen Erler, Ruihan Zhang, Loukas Petridis, Xiaolin Cheng, Jeremy~C Smith, and
  J{\"o}rg Langowski.
\newblock The role of histone tails in the nucleosome: a computational study.
\newblock {\em Biophysical journal}, 107(12):2911--2922, 2014.

\bibitem{li2016distinct}
Zhenhai Li and Hidetoshi Kono.
\newblock Distinct roles of histone h3 and h2a tails in nucleosome stability.
\newblock {\em Scientific reports}, 6:31437, 2016.

\bibitem{bignon2021dynamic}
Emmanuelle Bignon, Natacha Gillet, Tao Jiang, Christophe Morell, and Elise
  Dumont.
\newblock A dynamic view of the interaction of histone tails with clustered
  abasic sites in a nucleosome core particle.
\newblock {\em The Journal of Physical Chemistry Letters}, 12:6014--6019, 2021.

\bibitem{medina2021unraveling}
Exequiel Medina, Danielle~R Latham, and Hugo Sanabria.
\newblock Unraveling protein's structural dynamics: from configurational
  dynamics to ensemble switching guides functional mesoscale assemblies.
\newblock {\em Current opinion in structural biology}, 66:129--138, 2021.

\bibitem{bendandi2020role}
Artemi Bendandi, Alessandro~S Patelli, Alberto Diaspro, and Walter Rocchia.
\newblock The role of histone tails in nucleosome stability: An electrostatic
  perspective.
\newblock {\em Computational and Structural Biotechnology Journal},
  18:2799--2809, 2020.

\bibitem{huertas2021histone}
Jan Huertas, Hans~Robert Sch{\"o}ler, and Vlad Cojocaru.
\newblock Histone tails cooperate to control the breathing of genomic
  nucleosomes.
\newblock {\em PLoS computational biology}, 17(6):e1009013, 2021.

\bibitem{kameda2019histone}
Takeru Kameda, Akinori Awazu, and Yuichi Togashi.
\newblock Histone tail dynamics in partially disassembled nucleosomes during
  chromatin remodeling.
\newblock {\em Frontiers in molecular biosciences}, 6:133, 2019.

\bibitem{tunyasuvunakool2021highly}
Kathryn Tunyasuvunakool, Jonas Adler, Zachary Wu, Tim Green, Michal Zielinski,
  Augustin {\v{Z}}{\'\i}dek, Alex Bridgland, Andrew Cowie, Clemens Meyer, Agata
  Laydon, et~al.
\newblock Highly accurate protein structure prediction for the human proteome.
\newblock {\em Nature}, pages 1--9, 2021.

\bibitem{jumper2021highly}
John Jumper, Richard Evans, Alexander Pritzel, Tim Green, Michael Figurnov,
  Olaf Ronneberger, Kathryn Tunyasuvunakool, Russ Bates, Augustin
  {\v{Z}}{\'\i}dek, Anna Potapenko, et~al.
\newblock Highly accurate protein structure prediction with alphafold.
\newblock {\em Nature}, pages 1--11, 2021.

\bibitem{kameda2021free}
Takeru Kameda, Katsura Asano, and Yuichi Togashi.
\newblock Free energy landscape of rna binding dynamics in start codon
  recognition by eukaryotic ribosomal pre-initiation complex.
\newblock {\em PLOS Computational Biology}, 17(6):e1009068, 2021.

\end{thebibliography}
\end{document}